\documentstyle[aps,prl,multicol,epsfig]{revtex}
\draft

\begin{document}
\title{Conduction through a quantum dot near a singlet-triplet transition}
\author{M. Pustilnik$^{a,b}$ and L. I. Glazman$^{c}$\vspace{2mm}}
\address{$^{a}$Danish Institute of Fundamental Metrology, 
Anker Engelunds Vej 1, Building 307, Lyngby 2800, Denmark \\
$^{b}${\O}rsted Laboratory, Niels Bohr Institute, Universitetsparken 5,
Copenhagen 2100, Denmark \\
$^{c}$ Theoretical Physics Institute, University of Minnesota, Minneapolis,
MN 55455, USA}
\maketitle

\begin{abstract}
  Kondo effect in the vicinity of a singlet-triplet transition in a
  vertical quantum dot is considered. This system is shown to map onto
  a special version of the two-impurity Kondo model.  At any value of
  the control parameter, the system has a Fermi-liquid ground state.
  Explicit expressions for the linear conductance as a function of the
  control parameter and temperature $T$ are obtained. At $T=0$, the
  conductance reaches the unitary limit $\sim 4e^2/h$ at the triplet
  side of the transition, and decreases with the increasing distance
  to the transition at the singlet side. At finite temperature, the
  conductance exhibits a peak near the transition point.
\end{abstract}


\begin{multicols}{2}
The number of electrons $N$ in a Coulomb-blockaded dot is a
well-defined quantity at low temperatures. The existence of a finite
energy gap for excited states carrying a different charge, $e(N\pm
1)$, normally results in a suppressed low-temperature conductance
through the dot. However, the suppression may be largely
lifted\cite{exp1} if $N$ is odd. This phenomenon is explained in terms
of the Kondo effect\cite{classics}. In the case of tunneling through a
spin-degenerate state, the conductance may reach the unit quantum
$2e^2/h$, which corresponds to a perfect transmission\cite{AM} through
the dot. The origin of the perfect transmission is in the formation of
a ``Kondo cloud'' consisting of the itinerant electrons of the leads,
which tends to screen the spin of the dot. The formation of such a
collective state is accompanied by the appearance of the resonance for
electrons right at the Fermi energy. Of course, this effect takes
place only if the dot has a non-zero spin in the ground state. This is
always the case for odd $N$. If $N$ is even, then the spin of the dot
is typically zero. However, if the spacing $\delta$ between the last
doubly-occupied and the following empty one-electron states in the dot
is anomalously small, then the exchange interaction, according to the
Hund's rule, favors the triplet ground state of the
dot\cite{kinks}. For a quantum dot formed in a GaAs heterostructure,
$\delta$ can be tuned by means of a magnetic field. Indeed, because of
a very small electron effective mass, magnetic field applied
perpendicular to the plane of electron gas has a strong orbital
effect. At a certain critical value of $\delta$ the singlet-triplet
transition occurs. Tuning through such a transition was demonstrated
recently in the experiments on vertical quantum dots
\cite{Tarucha}. Similar effect also occurs in lateral
devices\cite{Weis}. The applied magnetic field causes also
spin splitting. However the Zeeman energy, which lifts the spin
degeneracy, may remain sufficiently small, even compared with the
Kondo temperature, because of a small value of the effective
$g$-factor. Therefore the Kondo effect persists throughout the entire
domain of parameters corresponding to the triplet spin state.

The singlet-triplet transition was addressed in the recent work
\cite{EN}, where quite special model of the quantum dot was
considered. As it can be shown, the additional symmetries assumed 
in\cite{EN} may result in a non-Fermi liquid behavior at the transition 
point.

As we show below, a generic model of a quantum dot undergoing the
singlet-triplet transition allows for a mapping onto a the 2-impurity
Kondo model.  Using this mapping, we found that at finite temperature,
the conductance $G$ is a nonmonotonic function of the control
parameter $K$ with an asymmetric peak near the transition point
$K=K^\ast$. The asymmetry becomes more pronounced at lower
temperatures, and $G(K)$ becomes constant($\sim 4 e^2/h$) at $T=0$ at
the triplet side of the transition ($K<K^\ast$), but decreases
monotonously with $K$ at $K>K^\ast$.  Despite the apparent
non-analyticity of $G(K)$, the system has a Fermi-liquid ground state
for all $K$, including the transition point. 

The magnetic properties of a quantum dot can be understood from the
following Hamiltonian:  
\begin{equation}
H_{\rm dot}=\sum_{ns}\epsilon _{n}d_{ns}^{\dagger }d_{ns}-E_{s}{\bf S}%
^{2}+E_{C}\left( N-{\cal N}\right) ^{2}.  \label{Hdot}
\end{equation}
Here, 
$N=\sum_{ns}d_{ns}^{\dagger }d_{ns}$ 
is the number of electrons in the dot, 
${\bf S=}\sum_{nss^{\prime }}d_{ns}^{\dagger }
\left( \overrightarrow{\sigma }_{ss^{\prime }}/2\right) 
d_{ns^{\prime }}$ 
is the dot's spin, and the parameters $E_{C}$ and $E_{s}$ are the charging and
exchange energies \cite{exchange}.  We restrict our attention to the
very middle of a Coulomb blockade valley with an even number of
electrons in the dot (the dimensionless gate voltage ${\cal N}$ is
tuned to an even integer value).  We assume that the spacing $\delta$
between the last filled and first empty orbital states is of the order
of $E_s$, and that $\delta$ is tunable, {\it e.g.}, by means of a
magnetic field $B$. In order to model the singlet-triplet transition in
the ground state of the dot, it is sufficient to consider these two
states; we will assign indices $n=\pm 1$ to them. The four low-energy
states of the dot can be classified according to their spin $S=0,1$
and it's $z$-projection $S^{z}$. Labeling the states by these two
quantum numbers, $\left| S,S^{z}\right\rangle $, we find:
\begin{eqnarray}
&&\left| 11\right\rangle =d_{+1\uparrow }^{\dagger }d_{-1\uparrow }^{\dagger
}\left| 0\right\rangle ,\;\left| 1-1\right\rangle =d_{+1\downarrow
}^{\dagger }d_{-1\downarrow }^{\dagger }\left| 0\right\rangle ,  \nonumber \\
&&\left| 10\right\rangle =\frac{1}{\sqrt{2}}\left( d_{+1\uparrow }^{\dagger
}d_{-1\downarrow }^{\dagger }+d_{+1\downarrow }^{\dagger }d_{-1\uparrow
}^{\dagger }\right) \left| 0\right\rangle ,  \label{Basis} \\
&&\left| 00\right\rangle =d_{-1\uparrow }^{\dagger }d_{-1\downarrow
}^{\dagger }\left| 0\right\rangle   \nonumber
\end{eqnarray}
where $\left| 0\right\rangle $ is the ground state of the dot with
${\cal N}-2$ electrons. Projected onto these states, the Hamiltonian
of the dot becomes (up to a constant)
\begin{equation}
{\cal P}H_{\rm dot}{\cal P}= 
\frac {K_{0}}{4}\sum_{S,S^z}
\left| S,S^z\right\rangle
\left (
\delta_{S,1}-3\delta_{S,0}
\right)
 \left\langle S,S^z\right|, 
\label{Hd}
\end{equation}
where $K_0=\delta -2E_{s}$ is the energy difference between the singlet 
and the triplet, and ${\cal P}$ is the projection operator on the
system of states (\ref{Basis}).

Upon the variation of magnetic field $B$, the singlet-triplet
transition occurs at $K_0 = 0$. Unlike the
special case considered in Ref.~\onlinecite{EN}, we are interested in
the generic model with $E_s\neq 0$ at the transition point.

In a vertical dot device, the potential creating lateral confinement
of electrons most probably does not vary much over the thickness of
the dot\cite{Tarucha}. Therefore we assume that the
electron orbital motion perpendicular to the axis of the device can be
characterized by the same quantum number $n$ inside the dot and in the
leads; this quantum number is conserved in the process of tunneling.
Thus, our model consists of the Hamiltonian of the dot, already
discussed above, the Hamiltonian of the leads
\begin{equation}
H_{l}=\sum_{\alpha nks}\xi _{k}c_{\alpha nks}^{\dagger }c_{\alpha nks},
\label{leads}
\end{equation}
and the tunneling Hamiltonian:  
\begin{equation}
H_{T}=\sum_{\alpha nks}t_{\alpha n}c_{\alpha nks}^{\dagger
  }d_{ns}+{\rm H.c.}.
\label{HT}
\end{equation}
Here $\alpha =R,L$ for the right/left lead, and $n=\pm 1$ for the two
orbitals participating in the singlet-triplet transition; $k$ labels
states of the continuum spectrum in the leads, and $s$ is the spin
index. After a rotation in the R-L space,
\[
\left( 
\begin{array}{c}
\psi _{nks} \\ 
\phi _{nks}
\end{array}
\right) = 
\frac{1}{t_n}
\left( 
\begin{array}{cc}
t_{Rn} & t_{Ln} \\ 
-t_{Ln} & t_{Rn}
\end{array}
\right) \left( 
\begin{array}{c}
c_{Rnks} \\ 
c_{Lnks}
\end{array}
\right) ,
\]
the $\phi$ field decouples: 
$H_{T}=\sum_{nks}t_{n}\psi_{nks}^{\dagger}d_{ns}+{\rm H.c.}$; here
$t_{n}^2=t_{Ln}^{2}+t_{Rn}^{2}$.
Next we integrate out the virtual transitions to the states with 
${\cal N}\pm 1$ electrons by means of the Schrieffer-Wolff transformation. 
The resulting effective low-energy Hamiltonian includes the operators 
\[
{\bf S}_{nn^{\prime }}={\cal P}\sum_{ss^{\prime }}d_{ns}^{\dagger }
\frac{\overrightarrow{\sigma }_{ss^{\prime }}}{2}
d_{n^{\prime }s^{\prime }}{\cal P}.
\]
The effective Hamiltonian may be conveniently written in terms of two
fictitious $1/2$-spins ${\bf S}_{1,2}$ which represent the same
symmetries as the set of states (\ref{Basis}). This one-to-one
correspondence between the basis states allows us to represent
operators ${\bf S}_{nn^{\prime }}$ in terms of ${\bf S}_{1,2}$. We find
the following relations:
\begin{eqnarray}
{\bf S}_{nn} &=&\frac{1}{2}\left( {\bf S}_{1}+{\bf S}_{2}\right) 
=\frac{1}{2}{\bf S}_{+},  
\nonumber \\
\sum_{n}{\bf S}_{-n,n} &=&\frac{1}{\sqrt{2}}\left( {\bf S}_{1}-{\bf S}
_{2}\right) =\frac{1}{\sqrt{2}}{\bf S}_{-},  
\label{new} \\
\sum_{n}in{\bf S}_{-n,n} &=&\sqrt{2}\left[ {\bf S}_{1}\times {\bf S}_{2}
\right] =\sqrt{2}{\bf T}.  
\nonumber
\end{eqnarray}
Using (\ref{new}), the effective Hamiltonian is written in a form,
resembling\cite{2IKM} the two-impurity Kondo model\cite{ALJ}: 
\begin{eqnarray}
H&=&\sum_{nks}\xi _{k}\psi _{nks}^{\dagger }\psi _{nks}+K\left( {\bf S}
_{1}\cdot {\bf S}_{2}\right) +\sum_{n}H_{n},  \label{model}\\
H_{n} &=&J_{n}\left( {\bf s}_{nn}\cdot {\bf S}_{+}\right) +Vn\rho
_{nn}\left( {\bf S}_{1}\cdot {\bf S}_{2}\right)   \label{Hn} \\
&&+\frac{I}{\sqrt{2}}\left[ \left( {\bf s}_{-n,n}\cdot {\bf S}_{-}\right)
+2in\left( {\bf s}_{-n,n}\cdot {\bf T}\right) \right].   \nonumber
\end{eqnarray}
Here the particle and  spin densities in the continuum are
\[
\rho _{nn}=\sum_{kk^{\prime }s}\psi _{kns}^{\dagger }\psi _{k^{\prime }ns},\;
{\bf s}_{nn^{\prime }}=\sum_{kk^{\prime }ss^{\prime }}\psi_{kns}^{\dagger }
\frac{\overrightarrow{\sigma }_{ss^{\prime }}}{2}\psi _{k^{\prime }n^{\prime
}s^{\prime }},
\]
and the bare values of the coupling constants are 
\begin{equation}
J_{n}=\frac{2t_{n}^{2}}{E_C},\;I=\frac{2t_{+1}t_{-1}}{E_C},\;
V=\frac{t_{+1}^{2}+t_{-1}^{2}}{2E_C}.  
\label{bare}
\end{equation}

We did not include in Eq.~(\ref{model}) the Hamiltonian of the $\phi$
field, and other terms which are irrelevant for the low-energy
renormalization. The Schrieffer-Wolff transformation also produces
corrections to $K_0$, so $K$ does not coincide with its bare value.
This difference is not important, as it only affects the position of
the singlet-triplet transition, but not its nature. A common factor $I$
in the last two terms of Eq.~(\ref{Hn}) comes from the conservation of
the orbital index $n$, see Eq.~(\ref{HT}).

To simplify the analysis of Eqs.~(\ref{model})-(\ref{bare}), we further
restrict our attention to the symmetric case $t_{\alpha n}=t_{\alpha }$,
for which the definition (\ref{bare}) reduces to 
\begin{equation}
J_n\equiv J=I=2V=2\left(t_L^2+t_R^2\right)/E_C
\label{equal}
\end{equation}
This simplification is adequate to the experimentally relevant case of
very thin barriers separating the dot from the leads, and, more
importantly, only insignificantly affects the low-energy properties of
the model.  To calculate the differential conductance in the leading
logarithmic approximation, we apply the ``poor man's'' scaling
technique \cite{PWA}. The procedure consists of a successive
integration out of the high-energy degrees of freedom, and yields the
set of scaling equations
\begin{equation}
\frac{dJ}{d{\cal L}} = \nu \left( J^{2}+I^{2}\right) ,  
\;
\frac{dI}{d{\cal L}} = 2\nu I\left( J+V\right) ,  
\;
\frac{dV}{d{\cal L}} = 2\nu I^{2},  
\label{Scaling}
\end{equation}
with the initial conditions (\ref{equal}). Here ${\cal L}=\ln E_C/D$,
and $\nu$ is the density of states in the leads; the initial value of
the energy cutoff $D$ is $D=E_C$. This procedure also generates
non-logarithmic corrections to $K$. In the following we absorb these
corrections in the re-defined value of $K$. Equations (\ref{Scaling})
are valid for $D\gg|K| ,T$. 

\begin{figure}[tbp]
\centerline{\epsfxsize=6.0cm
\epsfbox{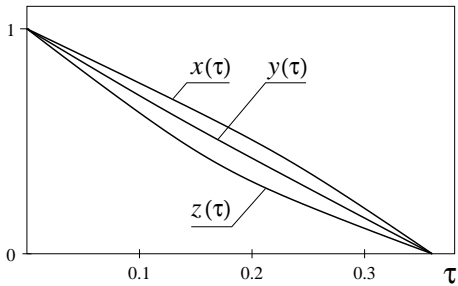}\vspace{1.5mm}}
\caption{ 
Numerical solution of the scaling equations. The RG equations (\ref{Scaling}) 
are rewritten in terms of the new variable $\tau =\nu J(0)\ln {E_C}/D$
and the new functions 
$x(\tau )=J(0)/J(\tau )$, 
$y(\tau )=I(0)/I(\tau )$, 
$z(\tau )=V(0)/V(\tau )$. 
The three functions reach zero simultaneously at $\tau =\tau_{0}=0.36$. 
}
\label{FigScaling}
\end{figure}

At certain value of ${\cal L}= {\cal L}_{0}$, the inverse coupling
constants simultaneously reach zero:
\[
1/J\left( {\cal L}_{0}\right) =1/I\left( {\cal L}_{0}\right) 
=1/V\left({\cal L}_{0}\right) =0. 
\]
This defines, through the equation ${\cal L}_{0}=\ln E_{C}/T_{0}$, the
characteristic energy scale of the problem:
\[
T_{0}=E_{c}\exp \left[ -\tau _{0}/\nu J(0)\right]. 
\]
Here $\tau _0$ is a parameter that depends on the initial conditions and should 
be found numerically.  We obtained $\tau _{0}=0.36$ (see Fig.\ref{FigScaling}).  
Thus, the strong coupling regime is reached at much higher temperatures, than in
the usual Kondo model (for which the Kondo temperature would be given
by the same expression as $T_{0}$, but with $\tau _{0}=1$).

The solution of the renormalization group (RG) equations (\ref{Scaling}) can 
now be expanded near 
${\cal L}={\cal L}_{0}$. To the first order in 
${\cal L}_{0}-{\cal L}=\ln D/T_{0}$, 
we obtain 
\begin{equation}
\frac{1}{\nu J}=\frac{\sqrt{\lambda }}{\nu I}=\frac{\lambda -1}{2\nu V}
=\left( \lambda +1\right) \ln D/T_{0},  \label{ScalingSolution}
\end{equation}
where $\lambda =2+\sqrt{5}\approx 4.2$ is a model-independent
constant, {\it i.e.}, it does not change if the restriction 
$t_{\alpha n}=t_{\alpha}$ is lifted.

The solution (\ref{ScalingSolution}) can be used to calculate the
differential conductance at high temperature $T\gg \left| K\right|
,T_{0}$.  In this regime, the coupling constants are still small, and
the conductance is obtained by applying a perturbation theory to the
Hamiltonian (\ref{model})-(\ref{Hn}) with renormalized parameters.
This yields
\begin{equation}
G/G_0=A \ln ^{-2}T/T_{0},
\label{transition}
\end{equation}
where 
\[
A=\left( 3\pi ^{2}/8\right) \left( \lambda +1\right) ^{-2}
\left[1+\lambda +\left( \lambda -1\right) ^{2}/8\right] 
\approx 0.9
\]
is a numerical constant, and
\begin{equation}
G_{0}=\frac{4e^2}{h}
\left ( \frac{2 t_L t_R}{t^2_L + t^2_R} \right )^2 .
\label{G0}
\end{equation}
Note that (\ref{transition}) includes contributions from both  the processes 
conserving the orbital index [the first two terms in Eq. (\ref{Hn})], and the 
processes involving an inter-orbital scattering. 

Away from the singlet-triplet degeneracy point, $|K|\gtrsim T_0$, the
RG flow yielding Eq.~(\ref{transition}) terminates at energy $D\sim
|K|$. At the triplet side of the transition ($K<0$), the two spins
${\bf S} _{1,2}$ are locked into a triplet state. The system is
described by the effective 2-channel Kondo model with $S=1$ impurity,
obtained from Eqs.~(\ref {model})-(\ref{Hn}) by projecting out the
singlet state:
\[
H_{{\rm triplet}}=\sum_{nks}\xi _{k}\psi _{nks}^{\dagger }\psi
_{nks}+J_{t}\sum_{n}\left( {\bf s}_{nn}\cdot {\bf S}\right)
+V_{t}\sum_{n}n\rho _{nn};
\]
here $J_{t}=J\left(|K|\right) ,\;V_{t}=V\left(|K|\right) /4$. As $D$ is
further lowered, $J_t$ is governed by the standard Kondo RG equation 
\[
dJ_{t}/d{\cal L}=\nu J_{t}^{2},\;{\cal L}=\ln D/\left| K\right| .
\]
The solution of this equation, $1/\nu J_{t}(D)=\ln D/T_{k}$, is expressed
through the $K$-dependent Kondo temperature $T_{k}\left( K\right) =\left|
K\right| \exp \left( -1/\nu J_{t}\right) $, which, using (\ref
{ScalingSolution}), is in turn expressed through $T_{0}$ as 
\begin{equation}
T_{k}/T_{0}=\left( T_{0}/\left| K\right| \right) ^{\lambda }.  \label{Tk}
\end{equation}
Recall that the exponent here, $\lambda \approx 4.2$, is universal. 
Eq.~(\ref{Tk}) was obtained also in \cite{EN}, but with a rather different 
value of the exponent $\lambda$ (according to \cite{EN}, 
$\lambda \leq 1$ and appears to be non-universal).

\begin{figure}[tbp]
\centerline{\epsfxsize=6.0cm
\epsfbox{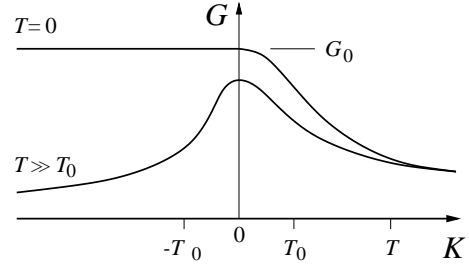}\vspace{1.5mm}}  
\caption{
$K$-dependence of the differential conductance at two different
temperatures. The two asymptotes merge at $K\gg T,T_0$.
}
\label{overall}
\end{figure}

Since $V_{t}$ is not renormalized, the differential conductance at
$T\ll -K$ is dominated by the exchange (the second term in $H_{\rm
  triplet}$) and is given by
\begin{equation}
G/G_{0}=f\left( {T}/{T_{k}}\right) =f\left[ \frac{T}{T_{0}}\left
(\frac{\left| K\right| }{T_{0}}\right) ^{\lambda }\right] ,
\label{triplet}
\end{equation}
where $f\left( {x}\right) $ is a smooth function that interpolates
between $f\left( x\gg 1\right) =\left( \pi ^{2}/2\right) \ln ^{-2}x$
and $f\left( 0\right) =1$. It coincides with the scaled resistivity
$f(T/T_K)=\rho (T/T_K)/\rho(0)$ for the symmetric two channel $S=1$
Kondo model.  The conductance at $T=0$ (the unitary limit value),
$G_{0}$, is given above in Eq.~(\ref{G0}), see also Fig.~\ref{overall}.

On the singlet side of the transition, $K\gtrsim T_0$, the scaling
terminates at $D\sim K$, and the low-energy effective Hamiltonian is
\[
H_{{\rm singlet}}=\sum_{nks}\xi _{k}\psi _{nks}^{\dagger }\psi
_{nks}+V_{s}\sum_{n}n\rho _{nn}, 
\]
where $V_{s}=-3V\left( K\right) /4$. The conductance at temperatures
$T\ll K$ is given by
\begin{equation}
G/G_0=B\ln ^{-2}K/T_0,
\quad 
B=\left( \frac{3\pi }{8}\frac{\lambda -1}{\lambda +1}\right)^2
\approx 0.5 .
\label{singlet}
\end{equation}

The only regime of parameters left beyond the range of the validity of
the equations (\ref{transition}), (\ref{triplet}) and (\ref{singlet})
is that of the very vicinity of the singlet-triplet transition $K=0$
at low temperature $T\lesssim T_{0}$. Understanding of this regime
requires knowledge of the properties of the system's ground state.
This can be inferred from the following simple symmetry-based
argument, devised originally for the two impurity Kondo model
\cite{ALJ}.  The Hamiltonian (\ref{model}), (\ref{Hn}) in the 
symmetric case $J_{n}=J$  is
invariant with respect to the particle-hole transformation
\begin{equation}
\psi _{n,k,s}\rightarrow s\psi _{-n,-k,-s}^{\dagger }.  \label{PH}
\end{equation}
Consider now the scattering problem at the Fermi energy: 
\begin{equation}
\Phi _{ns}^{\rm out}={\rm S}_{ns,n^{\prime }s^{\prime }}
\Phi _{n^{\prime}s^{\prime }}^{\rm in}.  
\label{scattering}
\end{equation}
Particle-hole symmetry (\ref{PH}) implies that for any $\Phi ^{\rm
  in(out)}$ which solves Eq.(\ref{scattering}), $\Phi ^{\rm
  in(out)}=i\Sigma \Phi ^{\rm out(in)}$ will be a solution as well
(here $\Sigma_{ns,n^{\prime }s^{\prime }} =
\sigma_{ss^{\prime}}^{y}\sigma_{nn^{\prime}}^{x}$, and $\sigma ^{i}$
are the Pauli matrices).  Substituting this into (\ref{scattering})
one finds a relation for the ${\rm S}$-matrix: $\left( \Sigma {\rm
    S}\right)^{2}=1$. Among the diagonal ${\rm S}$, this is satisfied
by ${\rm S}_{ns,n^{\prime}s^{\prime }}
=\delta_{nn^{\prime}}\delta_{ss^{\prime}}e^{-2in\theta}$ with
arbitrary $\theta $.  In other words, the particle-hole symmetry
imposes a restriction on the scattering phase shifts at the Fermi
energy: $\theta _{ns}= n\theta$.  The phase $\theta$, and therefore
scattering phases $\theta _{ns}$ vary {\it continuously} from
$\pm\pi/2$ to $0$ as $K$ is varied from $K=-\infty$ to $K=+\infty$.
At the triplet side of the transition, $-K\gg T_{0}$, the
low-temperature physics is described by the effective 2-channel $S=1$
Kondo model, as discussed above. In that model, the zero-energy phase
shift $\theta \equiv \pi /2$. This means that there exists $K^{\ast
  }\gtrsim -T_0$ such that $\theta (K)=\pi /2$ for all $K<K^{\ast }$.
On the other hand, at the singlet side of the transition, $\theta (K)$
decreases with $K$ logarithmically at large positive $K\gg T_{0}$, but
approaches $\pi/2$ at smaller $K$ because of larger values of $V_{s}$
in $H_{\rm singlet}$.  We expect therefore that $\theta (K)$ is a
continuous, though a non-analytical, function. The singlet-triplet
transition may be associated with the point of the non-analyticity
$K=K^\ast$.  According to these arguments, it is natural to conjecture
that the Fermi liquid description \cite{N} is applicable at all $K$
and thus the conductance at $T=0$ is $G=G_{0}\sin ^{2}\theta (K)$. The
inter-orbital scattering processes make no contribution to $G$ in this
regime, unlike at high temperature $T\gg |K|,T_{0}$.

It should be noticed that the problem under consideration is quite
different from what one faces if the Zeeman splitting is the leading
effect of the magnetic field \cite{Zeeman} (which may happen if
magnetic field is applied in the plane of the lateral dot). In
this case, the two-fold degeneracy of the ground state of an isolated
dot appears only at one special value of magnetic field, such that the
Zeeman energy equals $\delta$.  Accordingly, $G\left(T,B\right)$
exhibits a Kondo peak at this strength of the field at all
temperatures.

To conclude, we considered the linear conductance $G(T,K)$ of a
vertical quantum dot near the singlet-triplet transition. The
transition occurs due to the strong orbital effect of an external
magnetic field.  At high temperature $G$ exhibits a peak near the
transition point, in agreement with the experiments\cite{Tarucha}. 
At low temperature $G$ reaches the unitary
limit value at the triplet side of the transition, and decreases
monotonously at the singlet side (see Fig. \ref{overall}). The
characteristic energy scale of the effect is much larger than that for
the usual Kondo effect with the similar system parameters.

The authors are grateful to M. Eto, L. Kouwenhoven and J. Weis for
discussions. This work is supported by NSF under Grants
DMR-9812340, DMR-9731756 and by the European Commission through the
contract SMT4-CT98-9030. MP gratefully acknowledges the kind
hospitality of the Theoretical Physics Institute of the University of
Minnesota.

\end{multicols}

\end{document}